# Pressure-induced superconductivity in the three-dimensional topological Dirac semimetal $Cd_3As_2$

L. P. He[1,*], Y. T. Jia[2,*], S. J. Zhang[2,†], X. C. Hong[1], C. Q. Jin[2,4,†], and S. Y. Li[1,3,‡]

[1]*State Key Laboratory of Surface Physics, Department of Physics, and Laboratory of Advanced Materials, Fudan University, Shanghai 200433, P. R. China*
[2]*Beijing National Laboratory for Condensed Matter Physics and Institute of Physics, Chinese Academy of Sciences, Beijing 100190, P. R. China*
[3]*Collaborative Innovation Center of Advanced Microstructures, Fudan University, Shanghai 200433, P. R. China*
[4]*Collaborative Innovation Center of Quantum Matter, Beijing 100871, P. R. China*
*These authors contributed equally to this work.

**The recently discovered Dirac and Weyl semimetals are new members of topological materials. Starting from them, topological superconductivity may be achieved, e.g. by carrier doping or applying pressure. Here we report high-pressure resistance and X-ray diffraction study of the three-dimensional topological Dirac semimetal $Cd_3As_2$. Superconductivity with $T_c \approx 2.0$ K is observed at 8.5 GPa. The $T_c$ keeps increasing to about 4.0 K at 21.3 GPa, then shows a nearly constant pressure dependence up to the highest pressure 50.9 GPa. The X-ray diffraction measurements reveal a structure phase transition around 3.5 GPa. Our observation of superconductivity in pressurized topological Dirac semimetal $Cd_3As_2$ provides a new candidate for topological superconductor, as argued in a recent point contact study and a theoretical work.**

In recent few years, the search for topological superconductors (TSCs) has been a hot topic in condensed matter physics [1,2]. The TSCs have a full pairing gap in the bulk and gapless surface states consisting of Majorana fermions [1]. This is in close analogy to the topological insulators (TIs), which have a full insulating gap in the bulk and gapless edge or surface states [1]. The TSC is of great importance, since it is not only a new kind of exotic superconductor, but also one source of Majorana ferimons for future applications in quantum computations [1,2].

Experimentally, the simplest way to get a candidate for TSC is to convert a TI into superconductor, by tuning the parameters such as doping or pressure. For example, by doping, $Cu_xBi_2Se_3$ and $Cu_x(PbSe)_5(Bi_2Se_3)_6$ are considered to be candidates for TSCs [3-6], while $Sn_{1-x}In_xTe$ is considered as a candidate for topological crystalline superconductor [7,8]. Under pressure, $Bi_2Te_3$, $Bi_2Se_3$, $Sb_2Te_3$, and $Sb_2Se_3$ become superconducting, which are also regarded as candidates for TSCs [9-14]. Note that there are debates on whether these candidates are indeed TSCs [9-17], therefore further experimental works are needed to definitely identify a TSC and manipulate the Majorana fermions on its surface.

More recently, a new kind of topological material, the three-dimensional (3D) Dirac semimetal was discovered, with examples of $Na_3Bi$ and $Cd_3As_2$ [18-27]. As a 3D analogue to graphene, the Fermi surface of the 3D Dirac semimetal only consists of 3D Dirac points with linear energy dispersion in any momentum direction [18,22]. The exotic Fermi surface of $Na_3Bi$ and $Cd_3As_2$ was confirmed by the angle-resolved photoemission spectroscopy (ARPES) experiments [19-21, 23-25]. The compound $Cd_3As_2$ is of particular interests, since it is stable in air, unlike $Na_3Bi$. Based on quantum transport measurement, a nontrivial π Berry's phase is obtained, which provides bulk evidence for the existence of 3D Dirac semimetal phase in $Cd_3As_2$ [27]. By symmetry breaking, this 3D Dirac semimetal may be driven to a topological insulator or Weyl semimetal [22]. More interestingly, it was predicted that topological superconductivity may be achieved in $Cd_3As_2$ by carrier doping [22], but this has not been realized so far. Since pressure is an effective way to induce superconductivity in TIs [9-14], it will be very interesting to check whether superconductivity can be achieved by applying pressure on $Cd_3As_2$.

Here we present the resistance measurements on $Cd_3As_2$ single crystals under pressure up to 50.9 GPa. After an initial increase with pressure, the low-temperature resistance starts to decreases with pressure above 6.4 GPa. Superconductivity appears at 8.5 GPa with $T_c \approx 2.0$ K, and the $T_c$ increases to about 4.0 K at 21.3 GPa, then persists to the highest pressure 50.9 GPa. A structure phase transition around 3.5 GPa is also observed by X-ray diffraction measurements. These results suggest that $Cd_3As_2$ may be a new topological superconductor under high pressure.

**Results**

**Pressure-induced superconductivity.** Figure 1(a) shows the crystal structure of $Cd_3As_2$ [28]. The cubic Cd lattice with two vacancies resides in a face-centered cubic As lattice. Figure 1(b) plots a typical resistivity curve of $Cd_3As_2$ single crystal at 0 GPa. It is metallic and non-superconducting down to 1.5 K.

In Fig. 2, the resistance curves for $Cd_3As_2$ single crystal under various pressures are plotted. From Fig. 2(a), the temperature dependence of resistance already changes to insulating behavior ($dR/dT < 0$) at 1.1 GPa. With increasing pressure, it becomes more and more insulating until 6.4 GPa. However, upon further increasing pressure, the resistance at low temperature decreases with pressure. In Fig. 2(b), it becomes more and more metallic up to 32.7 GPa. Figure 2(c) and 2(d) show the low-temperature part of the resistance curves above 8.5 GPa. A drop of resistance is observed below 2.0 K at 8.5 GPa, which is like a superconducting transition. At 11.7 GPa, the resistance drops to zero, and the transition temperature $T_c$ = 3.3 K is defined at the cross of the two straight lines. The $T_c$ increases to about 4.0 K at 21.3 GPa, then persists to the highest pressure 50.9 GPa.

To make sure the resistance drop in Fig. 2 is a superconducting transition, we measure the low-temperature resistance under 13.5 GPa in magnetic fields applied perpendicular to the (112) plane, as shown in Fig. 3(a). The resistance drop is gradually suppressed to lower temperature with increasing field, which demonstrates that it is indeed a superconducting transition.

Figure 3(b) plots the temperature dependence of $H_{c2}$. Although limited by the temperature range we measured, one can see an apparently linear temperature dependence of $H_{c2}$. With a linear fit to the data, $H_{c2}(0) \approx 4.29$ T is roughly estimated. This value is higher than the orbital limiting field $H_{c2}^{orb}(0) = 0.72T_c|dH_{c2}/dT|_{T=Tc} = 3.71$ T, according to Werthamer-Helfand-Hohenberg (WHH) formula [29]. It is much lower than the Pauli limiting field $H_P(0) = 1.84T_c = 7.89$ T [30,31], suggesting an absence of Pauli pair breaking. The linear temperature dependence of $H_{c2}$ in Fig. 3(b) is actually very interesting. It may come from a two-band Fermi surface topology as in $MgB_2$ [33,34], or an unconventional superconducting state as in heavy-fermion compound $UBe_{13}$ [35]. Similar linear temperature dependence of $H_{c2}$ has recently been observed in pressurized TSC candidates $Bi_2Se_3$ and $Cu_xBi_2Se_3$, and in non-centrosymmetric superconductor YPtBi under ambient and high pressures, which was considered as an indication of unconventional superconducting state [11,36,37].

We notice that no superconductivity was observed up to 13.43 GPa in an earlier pressure study of $Cd_3As_2$ single crystal [38]. The reason may be that their sample is slightly different from ours, and pressure higher than 13.43 GPa is needed to induce superconductivity. Interestingly, we also notice two recent point contact studies on $Cd_3As_2$ polycrystal and single crystal, respectively [39,40]. In both studies, indication of superconductivity was found around the point contact region on the surface, with $T_c$ comparable to ours. In particular, no superconductivity is observed by the "soft" point contact technique, therefore it was suggested that the superconductivity observed around the point contact region under the "hard" tip might be induced by the local pressure [40]. In this sense, our bulk resistance measurements under hydrostatic pressure confirm pressure-induced superconductivity in $Cd_3As_2$.

**Pressure-induced crystal structure phase transition.** Before discussing whether the pressure-induced superconductivity is topological or not, it is important to know whether it is accompanied by a structure phase transition, as observed in pressurized TIs [9-14]. High-pressure powder XRD measurements on $Cd_3As_2$ were performed up to 17.80 GPa. In Fig. 4, the XRD patterns below 2.60 GPa can be well indexed as the tetragonal phase in space group $I4_1/acd$ [28]. All the peaks are slightly shifted to higher angle with increasing pressure, due to the shrink of the lattice. However, when the pressure increases to 4.67 GPa and above, a set of new peaks emerges which is clearly different from that of low-pressure tetragonal phase. This abrupt change indicates that a new crystal structure phase appears, and we roughly determine the transition pressure around 3.5 GPa. Similar high-pressure XRD patterns have been observed in an earlier work, and the new high-pressure phase was determined as monoclinic in space group $P2_1/c$ [38].

**The unusual $T_c - p$ phase diagram.** In Fig. 5, we plot the temperature vs pressure phase diagram for $Cd_3As_2$. Since the resistance was only measured down to 1.8 K, we can not judge whether the superconductivity emerges at the same time as the structural transition near 3.5 GPa, or inside the high-pressure phase. Nevertheless, after increasing from 1.8 to about 4.0 K, there is apparently a region of constant $T_c$ from 21.3 to 50.9 GPa. Such a phase diagram is very similar to that of 3D TI $Bi_2Se_3$, which also shows a nearly constant $T_c$ from 30 to 50 GPa after an initial increase of $T_c$ starting from 12 GPa [11]. A constant $T_c$ over such a large pressure range is highly anomalous, as Kirshenbaum *et al.* already pointed out [11]. For $Bi_2Se_3$, two mechanisms with contrasting pressure-dependant $T_c$ may be balanced so as to produce a pressure-invariant $T_c$ over a wide range of pressure [11]. It was argued that the unique pressure evolution of $T_c$ and the anomalous linear temperature dependence of $H_{c2}$ are two evidences for unconventional superconductivity in $Bi_2Se_3$ [11]. The similarity between $Cd_3As_2$ and $Bi_2Se_3$ under pressure is worthy of further investigation.

**Discussion**

Now we discuss whether the superconducting state of $Cd_3As_2$ under high pressure is topological or not. In Ref. [40], the observation of zero bias conductance peak (ZBCP) and double conductance peaks (DCPs) under "hard" tip reveal *p*-wave like unconventional superconductivity in $Cd_3As_2$. Considering its special topological property, they suggested that $Cd_3As_2$ under high pressure is a candidate of the TSC [40]. Furthermore, a recent theoretical work also argued that $Cd_3As_2$ likely realizes a TSC with bulk point nodes and a surface Majorana fermion quartet [41]. Under high pressure, the symmetry lowering effect may stabilize the TSC phase by increasing the condensation energy, since the point nodes in the TSC phase are gapped when $C_4$ reduces to $C_2$ (the structure phase transition from tetragonal to monoclinic) [41]. These two works suggest that the superconductivity we observe under hydrostatic pressure is topological, although detailed band structure calculation for the high-pressure phase of $Cd_3As_2$ is needed to give more information about this possible TSC phase.

In summary, we have done resistance measurements on the 3D Dirac semimetal $Cd_3As_2$ single crystals under pressures up to 50.9 GPa. It is found that superconductivity with $T_c \approx 2$ K emerges at 8.5 GPa. The $T_c$ increases to 4.0 K at 21.3 GPa, then it shows an anomalous constant pressure dependence up to the highest pressure measured. High-pressure powder x-ray diffraction measurements reveal a structure phase transition around 3.5 GPa. Our observation of superconductivity in $Cd_3As_2$ under high pressure provides an interesting candidate for topological superconductor.

**Methods:**

High-quality $Cd_3As_2$ single crystals were grown from Cd flux [27]. The largest natural surface was determined as (112) plane by X-ray diffraction (XRD). The resistivity in vacuum (0 GPa) was measured on a large sample with dimension of 1.50 × 0.40 mm$^2$ in the (112) plane and 0.15 mm in thickness. The resistance measurement under pressure between 1.1 and 50.9 GPa was performed using diamond anvil cell (DAC) with solid transmitting medium hexagonal boron nitride (h-BN) [9,13,14]. The sample size is about 80 × 80 μm$^2$ in the (112) plane, with the thickness of ~10 μm. The pressure was determined by ruby fluorescence method at room temperature before and after each cooling down. The high-pressure powder XRD measurements with synchrotron radiation were performed at the HPCAT of Advanced Photon Source of Argonne National Lab using a symmetric Mao Bell diamond anvil cell at room temperature. The X-ray wavelength is 0.434 Å.

**Acknowledgements:** We thank X. Dai and Z. Fang for helpful discussions. This work is supported by the Ministry of Science and Technology of China (National Basic Research Program No. 2012CB821402 and 2015CB921401), the Natural Science Foundation of

**Author Contributions:** L.P.H. and X.C.H. grew the single crystals of $Cd_3As_2$. Y. T. J., S. J. Z. performed the transport measurements and analysed the data. L.P.H., and S.Y.L. wrote the manuscript. S.Y.L. and C. Q. J. supervised the project.

**Additional Information:** Correspondence and requests for materials should be addressed to S. J. Zhang (middleor@126.com), C. Q. Jin (jin@iphy.ac.cn) and S. Y. Li (shiyan_li@fudan.edu.cn).

**Competing financial interests:** The authors declare no competing financial interests.

**Figure 1 | Crystal structure and resistivity of $Cd_3As_2$.**

**(a)** The crystal structure of $Cd_3As_2$. The cubic Cd lattice with two vacancies resides in a face-centered cubic As lattice. **(b)** A typical resistivity curve of $Cd_3As_2$ single crystal at 0 GPa.

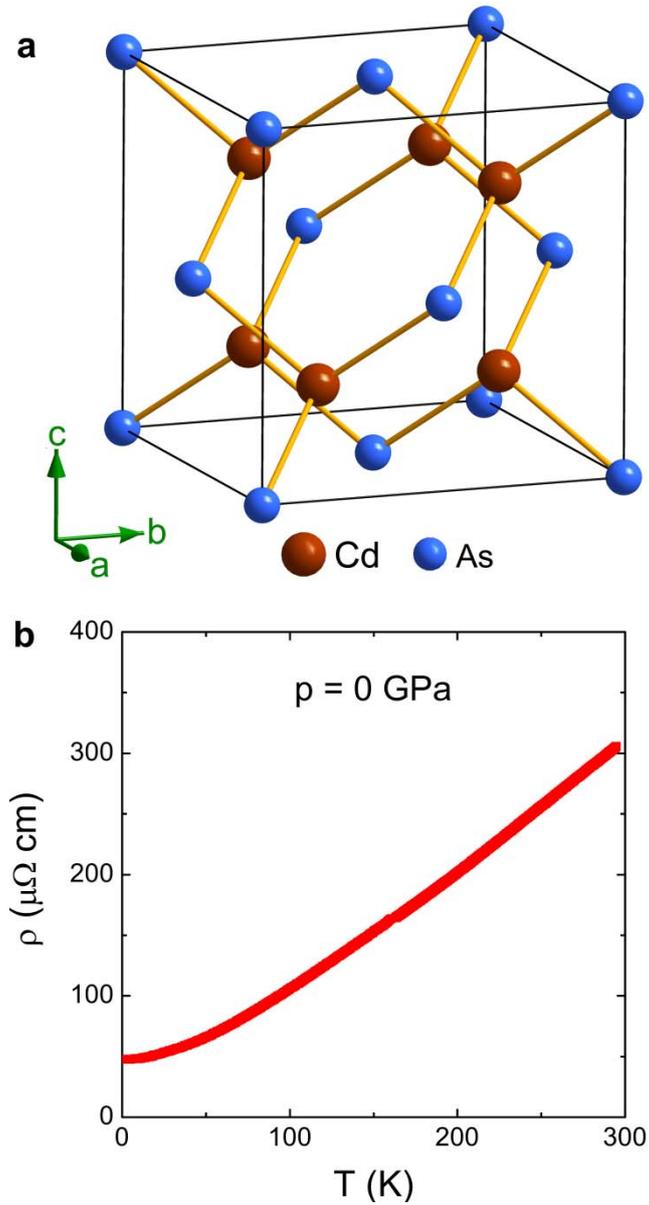

**Figure 2 | Experimental evidence of pressure-induced superconductivity.**

The temperature dependence of resistance for $Cd_3As_2$ single crystal under various pressures. **(a)** and **(b)**: the resistance from 1.8 to 300 K. **(c)** and **(d)**: low-temperature resistance showing the superconducting transition. The superconductivity appears at $p$ = 8.5 GPa with $T_c$ ≈ 2.0 K. The $T_c$ is defined as on the curve of $p$ = 11.7 GPa. The $T_c$ increases to about 4.0 K at 21.3 GPa, then persists to the highest pressure 50.9 GPa.

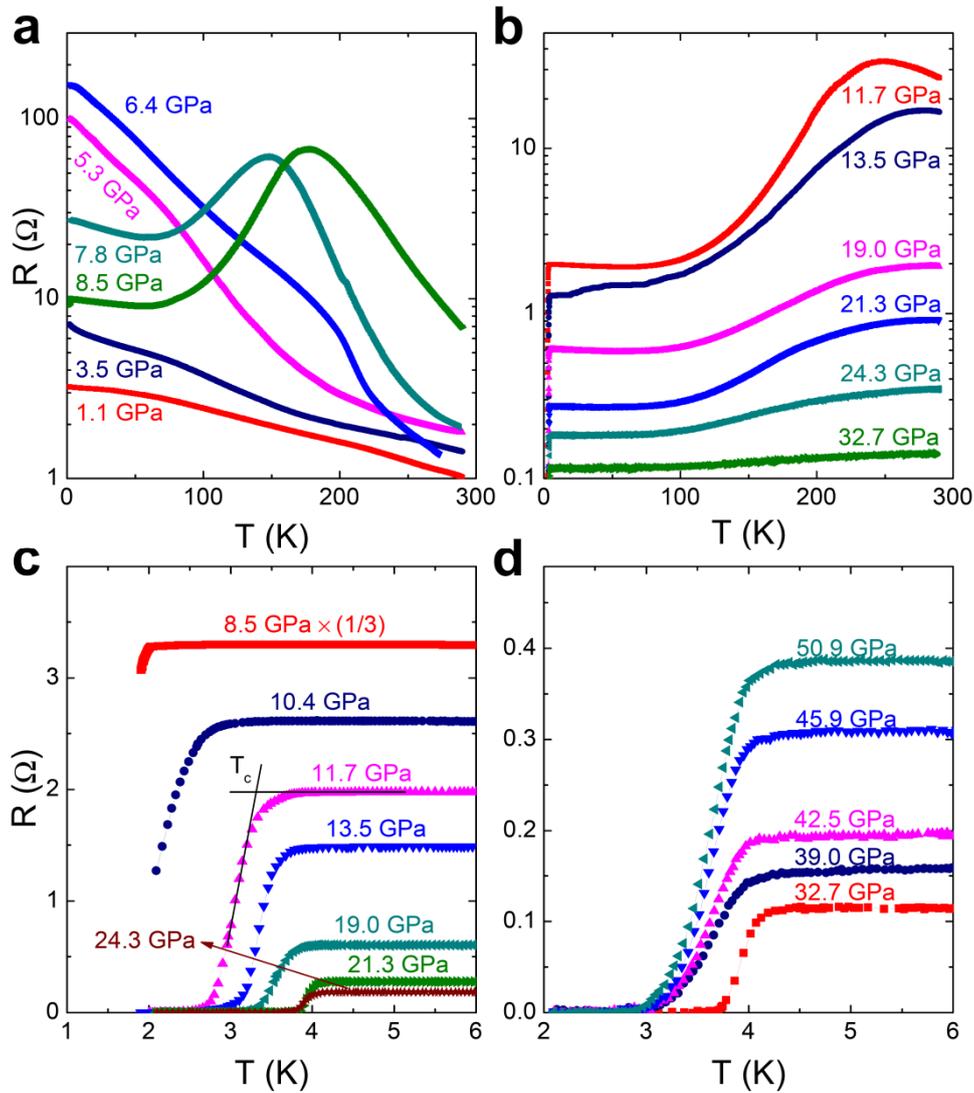

**Figure 3 | The upper critical field $H_{c2}$ of $Cd_3As_2$ under 13.5 GPa.**

**(a)** The superconducting transition of the $Cd_3As_2$ single crystal under 13.5 GPa and in magnetic fields applied perpendicular to the (112) plane. **(b)** Temperature dependence of the upper critical field $H_{c2}$. The dashed line is a linear fit to the data, which points to $H_{c2}$ (0) ≈ 4.29 T.

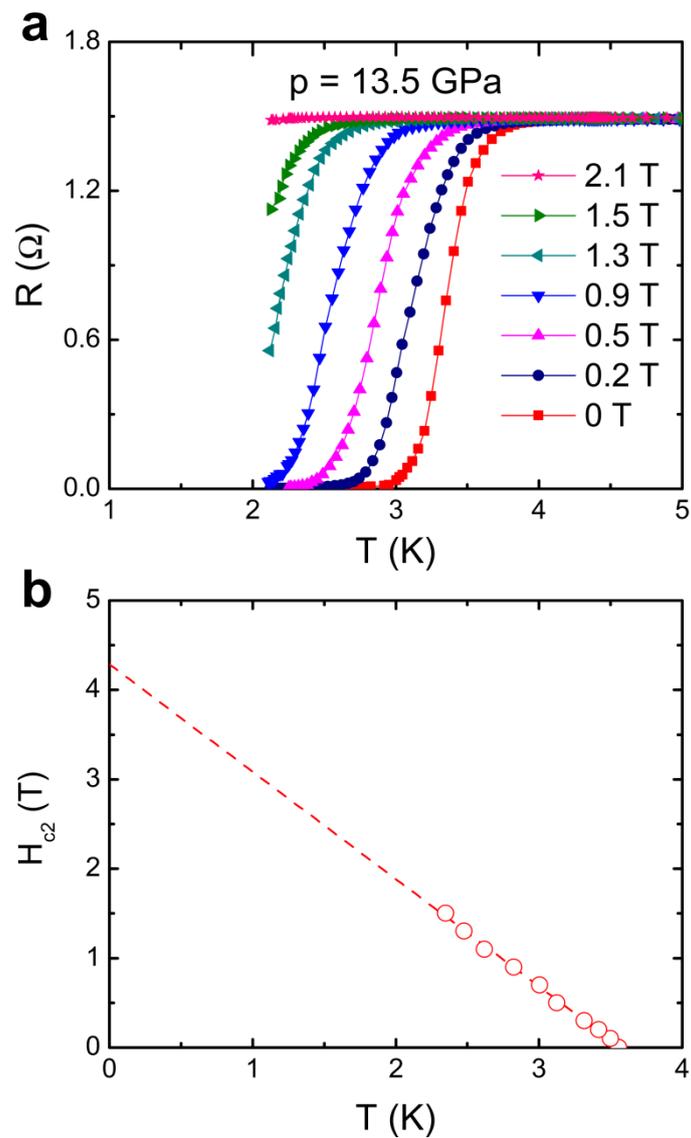

**Figure 4 | Crystal structure phase transition of $Cd_3As_2$ under pressure.**

The powder x-ray diffraction patterns of $Cd_3As_2$ under different pressures at room temperature. Below 2.60 GPa, the XRD patterns can be well indexed as the tetragonal phase in space group $I4_1/acd$ (shown by short black lines). A set of new peaks emerges when increasing pressure to 4.67 GPa and above, which shows a structure phase transition from tetragonal to monoclinic phase.

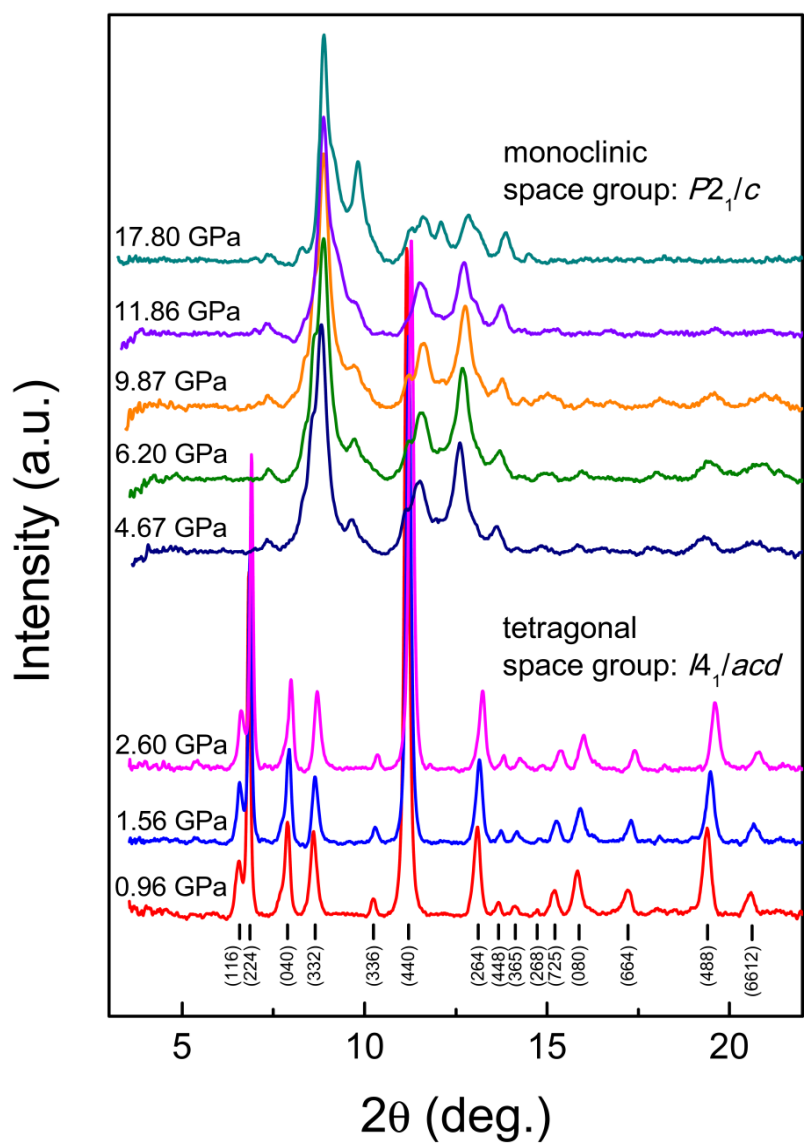

**Figure 5 | The phase diagram of $Cd_3As_2$.**

Temperature vs pressure phase diagram of $Cd_3As_2$. A structure phase transition occurs between 2.60 and 4.67 GPa. After increasing from 1.8 to about 4.0 K, there is apparently a region of constant $T_c$ from 21.3 to 50.9 GPa. Such a phase diagram is similar to that of the 3D topological insulator $Bi_2Se_3$.

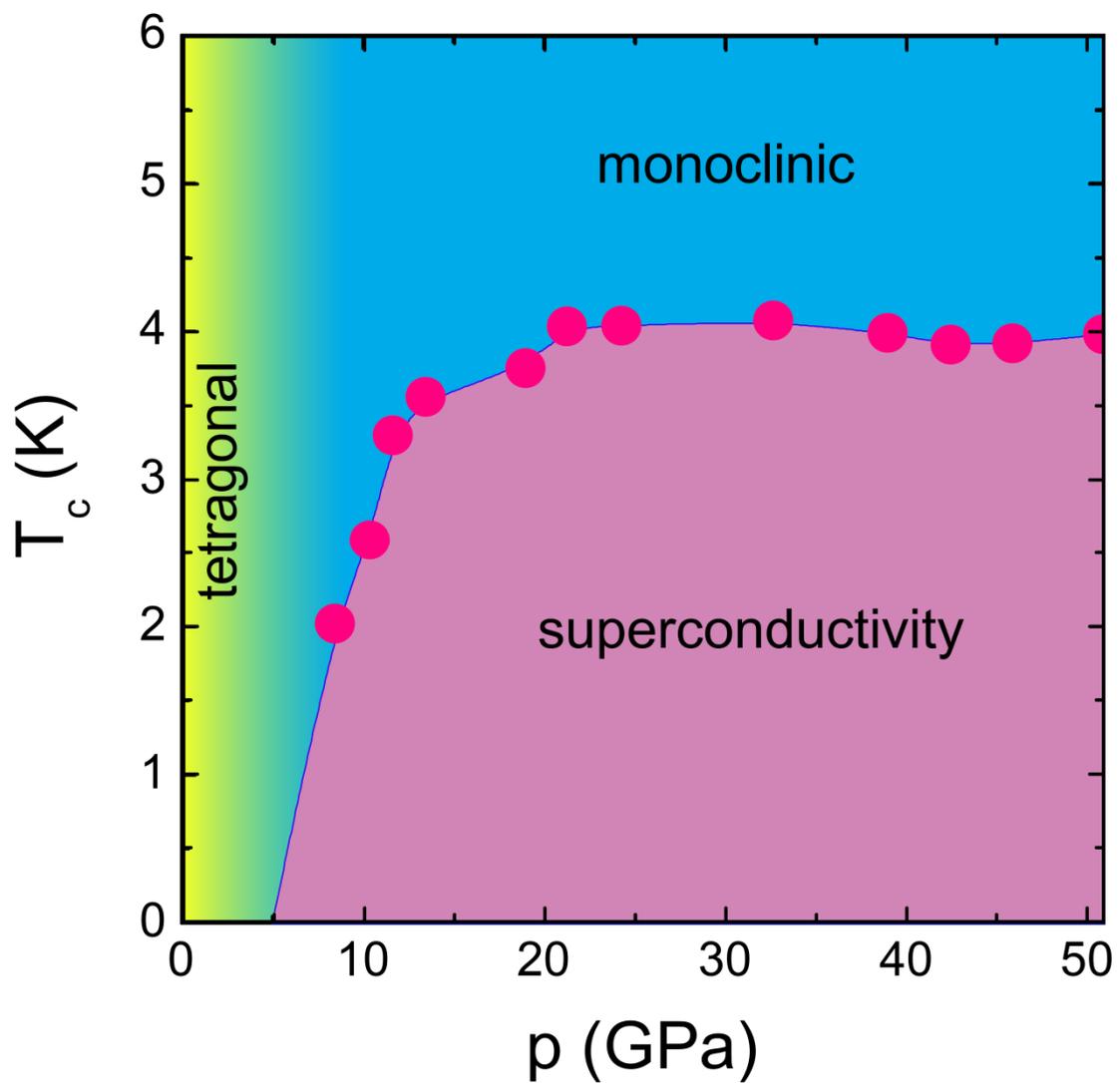